# Aliphatic Chains as One-Dimensional XY Spin Chains

Kirill Sheberstov*

*Chimie Physique et Chimie du Vivant (CPCV, UMR 8228), Department of Chemistry, Ecole Normale Supérieure, PSL University, Sorbonne University, CNRS, Paris, 75005, France*

23 December 2025

*correspondence to

kirill.sheberstov@ens.psl.eu

## Abstract

Spin waves are propagating disturbances of spin order in lattices with nearest-neighbor interactions. They are traditionally observed in magnetically ordered solids using inelastic neutron, light or electron scattering, and ferromagnetic resonance. Here we show that analogous spin dynamics can arise in liquid-state nuclear magnetic resonance (NMR) of molecules containing aliphatic chains. In such molecules, each $CH_2$ group must have a distinct chemical shift and be magnetically inequivalent via out-of-pair couplings. Under these conditions, singlet state populations of geminal protons propagate along $(CH_2)_n$ segments forming magnetically silent spin waves. For a chain with translational symmetry, the spin Hamiltonian factorizes into subspaces formally equivalent to the one-dimensional XY model. This correspondence yields analytic expressions for eigenstates and eigenenergies in a spectroscopy we term spin-chain zero-quantum NMR. We identify molecular systems in which these conditions are met. Their collective dynamics rapidly exceed classical computational tractability, making them targets for quantum-computer simulations of spin transport and many-body dynamics.

## Introduction

The modern understanding of collective spin dynamics traces back to early lattice descriptions of magnetism introduced by Lenz (*1*), developed into exactly solvable models by Ising (*2*), and placed on a quantum-mechanical foundation by Bloch through the discovery of collective spin-wave excitations (*3*). A closely related mathematical structure has been used to describe electronic bands in periodic crystals (*4*) and the formation of delocalized π molecular orbitals in conjugated systems (*5*), reflecting the universality of wave-like excitations in periodic quantum networks (*6*). Spin waves in solids are conventionally observed using inelastic neutron scattering (*7*), inelastic light scattering (*8*), spin-polarized electron energy-loss spectroscopy (*9*, *10*), and ferromagnetic resonance (*11*). These techniques provide detailed characterization of spin waves in crystalline materials, but their applicability is limited to the solid state.

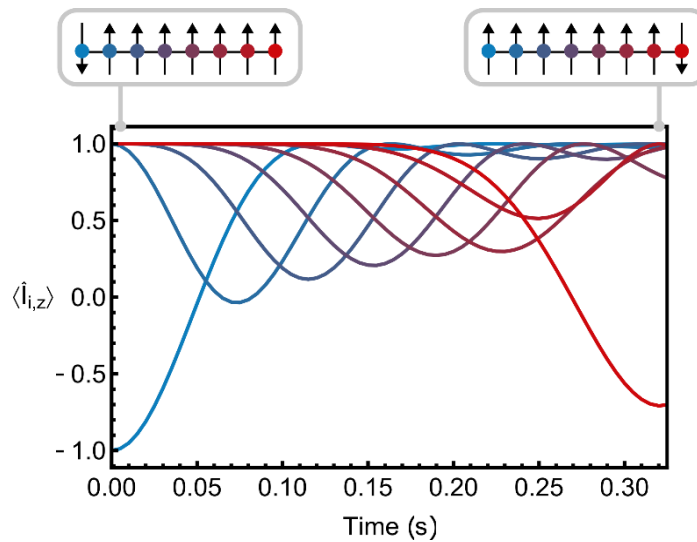

**Fig. 1**. **Propagation of a localized spin inversion in a linear chain of coupled spin-½ particles.** Simulated evolution of $\langle \hat{I}_{i,z}(t) \rangle$ expectation values for an eight-spin chain governed by the XY interaction (Eq. (1)) with a uniform scalar coupling of $J = 5$ Hz. An inversion initialized on the first spin propagates coherently along the chain, while the total $\hat{I}_z$ (eq. (2)) remains conserved. Individual spin polarizations oscillate in time, forming a traveling spin-polarization wave.

In contrast, the observation of spin waves in liquids has long been considered infeasible. Rapid molecular motion averages anisotropic spin–spin interactions, such as dipole–dipole couplings, therefore confining coherent dynamics to small groups of spins. Even NMR of macromolecules, such as liquid-state NMR of proteins containing tens of thousands of atoms, each amino acid forms an independent spin system. As a result, liquid-state NMR is typically described in terms of local spin dynamics rather than collective transport phenomena. Here, we challenge this prevailing view by demonstrating that collective spin dynamics can emerge in liquids in an unexpected form.

**Fig. 1** illustrates the type of collective dynamics central to this work. We consider the XY Hamiltonian arising from the non-secular part of the scalar $J$-coupling between neighboring spin-1/2 particles,

$$\hat{H}_{\text{XY}} = 2\pi J \sum_{i=1}^{n-1} (\hat{I}_{i,x}\hat{I}_{(i+1),x} + \hat{I}_{i,y}\hat{I}_{(i+1),y}). \tag{1}$$

Here, n denotes the total number of spins-1/2 in the chain, and $\hat{I}_{i,x/y}$ are the Cartesian components of the spin operator for site i. When a localized spin inversion is prepared at one end of the chain, the XY interaction redistributes the longitudinal spin polarization while conserving the total $z$-projection:

$$\hat{I}_z = \sum_i \hat{I}_{i,z}. \tag{2}$$

Numerical simulations for chains of varying length show that the time required for the wave to reach the opposite end increases linearly with chain length. While the total $z$-projection $\hat{I}_z$ is conserved, individual site polarizations $\hat{I}_{i,z}(t)$ oscillate in time, producing a traveling spin-polarization wave.

Here we show that in general case liquid-state NMR experiments on molecules containing aliphatic chains can exhibit analogous collective dynamics in high magnetic fields. In recent years, a progress has been made in uncovering experimentally accessible nuclear spin singlet states of protons in methylene groups of aliphatic chains containing two or three consecutive $CH_2$ units (*12–15*). In an isolated $CH_2$ group of an achiral molecule, the singlet state of the geminal protons ($|S_0\rangle = 1/\sqrt{2}\,[|\alpha\beta\rangle - |\beta\alpha\rangle]$) remains disconnected from the triplet states ($|T_{+1}\rangle = |\alpha\alpha\rangle$, $|T_0\rangle = 1/\sqrt{2}\,[|\alpha\beta\rangle + |\beta\alpha\rangle]$, and $|T_{-1}\rangle = |\beta\beta\rangle$), but in aliphatic chains the out-of-pair vicinal couplings between adjacent $CH_2$ groups may break this symmetry and render the geminal protons magnetically inequivalent (*16*). Magnetic inequivalence enables controlled creation of imbalances between the population of the singlet state and the mean population of the triplet states (*12, 13, 17–20*).

When the populations of the local singlet states are prepared in a non-uniform pattern along the chain, the resulting imbalance propagates through the scalar-coupling network in a manner closely related to the dynamics shown in **Fig. 1**. Experimentally, such transport of singlet order has been demonstrated for *chiral* molecules containing two $CH_2$ groups by Devience *et al.* (*21*), and later extended to achiral systems with three $CH_2$ units (*22*). As we show here, these observations are not isolated phenomena but arise from a deeper structural correspondence. Each $CH_2$ group possesses a two-level subspace spanned by the singlet $|S_0\rangle$ and central triplet $|T_0\rangle$ states, which can be viewed as a fictitious spin-½ particle. As we show below, the scalar $J$-coupling Hamiltonian that links neighboring $CH_2$ groups in aliphatic chains is reminiscent to the XY Hamiltonian introduced in Eq. (1). This similarity of Hamiltonians leads to analogous collective dynamics, including the propagation of singlet–triplet imbalances as a one-dimensional spin wave for arbitrary number of CH$_2$ groups in the molecule.

Recent developments in quantum simulation methods have highlighted the importance of physical systems that naturally generate complex, many-body spin dynamics, which are difficult to model classically (*23*, *24*). The spin wave propagation uncovered here in aliphatic chains provides a simple and experimentally accessible example of such behavior, enabling coherent transport to be observed with high spectral resolution in standard liquid-state NMR. These properties make CH$_2$ chains a promising platform for benchmarking quantum-simulation methods and for exploring quantum transport phenomena relevant to emerging quantum technologies.

## Theory

*Magnetic inequivalence of methylene groups in aliphatic chains*

Consider first an achiral molecule containing an isolated CH$_2$ group. In a strong external magnetic field, the equilibrium spin polarization of the two geminal protons is described (in the high-temperature approximation) by a density operator proportional to $\hat{I}_{1,z} + \hat{I}_{2,z}$, which corresponds to an overpopulation of the triplet state $|T_{+1}\rangle$ and a depopulation of $|T_{-1}\rangle$, while the populations of $|T_0\rangle$ and the singlet $|S_0\rangle$ remain equal. The three triplet states are symmetric under exchange of the two protons, whereas the singlet state is antisymmetric. For any Hamiltonian that is symmetric under intrapair permutation, unitary evolution cannot mix the singlet state with the triplet manifold, so the singlet remains disconnected from the triplets (*25*).

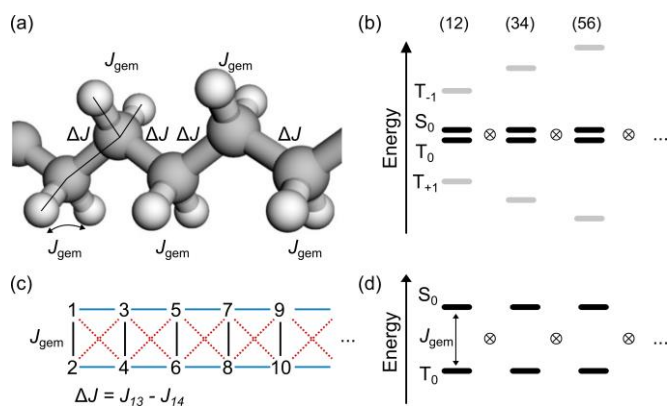

**Fig. 2. Spin topology of an aliphatic chain.** (a) Molecular structure of a representative aliphatic segment, highlighting the dominant intrapair geminal couplings $J_{\text{gem}}$ and the out-of-pair vicinal couplings to neighboring CH$_2$ groups. (b) Energy-level structure of an isolated CH$_2$ group, consisting of three triplet states and one singlet state; multiple such units appear along the chain. (c) Spin-network representation showing the numbering of spins and the pattern of couplings repeating with translational symmetry. Black lines denote intrapair geminal couplings $J_{\text{gem}}$. Blue solid lines represent gauche vicinal couplings ($J_{13} = J_{24} = J_{35} = \cdots$), while red dashed lines represent anti vicinal couplings ($J_{14} = J_{23} = J_{36} = \cdots$). The difference between gauche and anti couplings, $\Delta J = J_{\text{gauche}} - J_{\text{anti}}$, breaks intrapair permutation symmetry and renders the geminal protons magnetically inequivalent. (d) Effective two-level subsystems formed by the singlet $|S_0\rangle$ and central triplet $|T_0\rangle$ states of each CH$_2$ group, which govern the zero-quantum dynamics studied in this work. The singlet–triplet splitting is dominated by $J_{\text{gem}}$, while the outer triplet states $T_{\pm 1}$ are separated by Zeeman interactions and do not participate directly in the zero-quantum dynamics.

The situation changes in achiral molecules that contain an aliphatic chain of n chemically inequivalent CH$_2$ groups (**Fig. 2**). In such chains, each methylene unit is coupled to its neighbors through vicinal scalar couplings. When the rotationally averaged out-of-pair couplings of the two geminal protons to spins in a neighboring CH$_2$ group are unequal, the intrapair permutation symmetry is broken and the two protons become magnetically inequivalent (*16*, *26*). For example, if spins 1 and 2 form one CH$_2$ group and spins 3 and 4 belong to the adjacent group, magnetic inequivalence arises whenever $J_{13} \neq J_{14}$, where $J_{ij}$ denotes the scalar coupling between spins i and j. The strength of magnetic inequivalence is quantified by the difference:

$$\Delta J = J_{13} - J_{14} = J_{24} - J_{23}. \tag{3}$$

In practical terms, whenever $\Delta J \neq 0$, NMR radiofrequency (RF) pulses can be used to selectively manipulate the populations of the singlet and triplet states of a particular CH$_2$ group (*12*, *13*). The magnetic inequivalence is typically small, with $|\Delta J| \ll |J_{\text{gem}}|$, so that the large geminal coupling energetically separates the singlet from the triplet state. As a result, singlet and triplet states remain approximate eigenstates of the spin Hamiltonian, a situation often referred to as "singlet protection" (*17*). Despite this energetic protection, controlled transitions between the singlet and triplet manifolds can be deliberately induced using tailored RF sequences. In particular, the spin-lock induced crossing (SLIC) approach (*12*, *22*, *27*) enables selective preparation of non-equilibrium singlet–triplet population imbalances within individual CH$_2$ units.

*The spin Hamiltonian of (CH$_2$)$_n$ protons and its relation to a one-dimensional XY spin chain*

**Fig. 2** illustrates an aliphatic chain containing n consecutive CH$_2$ groups, together with the relevant spin–spin couplings and the energy-level structure of individual CH$_2$ units. In the following, we consider an idealized model in which the pattern of intrapair and vicinal scalar couplings obeys translational symmetry along the chain. Each CH$_2$ group is further assumed to possess a distinct chemical shift, such that the resonance-frequency differences between neighboring groups are much larger than the corresponding vicinal couplings. Under these weak-coupling conditions, adjacent CH$_2$ units are only weakly mixed by scalar couplings, so that the inter-pair scalar couplings can be treated in the secular (zz) approximation.

Writing the two vicinal couplings between adjacent CH$_2$ groups as $J_{gauche}$ (gauche) and $J_{anti}$ (anti), which correspond to $J_{13}$ and $J_{14}$ in eq. (3) under the assumption of translational symmetry, we define their symmetric and antisymmetric combinations as $\Sigma J = J_{gauche} + J_{anti}$, $\Delta J = J_{gauche} - J_{anti}$. This definition of $\Delta J$ is identical to that introduced in eq. (3), but is now applied uniformly along the chain. Using this notation, the *J*-coupling Hamiltonian for an aliphatic chain becomes:

$$\hat{H}_J = 2\pi J_{gem} \sum_{i=1}^{n} \hat{\mathbf{I}}_{2i-1} \cdot \hat{\mathbf{I}}_{2i} + 2\pi \frac{\Sigma J}{2} \sum_{i=1}^{n-1} (\hat{I}_{(2i-1),z} + \hat{I}_{2i,z})(\hat{I}_{(2i+1),z} + \hat{I}_{(2i+2),z}) + 2\pi \frac{\Delta J}{2} \sum_{i=1}^{n-1} (\hat{I}_{(2i-1),z} - \hat{I}_{2i,z})(\hat{I}_{(2i+1),z} - \hat{I}_{(2i+2),z}). \tag{4}$$

Here $J_{gem} = J_{12} = J_{34} = \ldots$ is geminal *J*-coupling between the two protons attached to the same carbon; n denotes the total number of CH$_2$ groups in the aliphatic system; $\hat{\mathbf{I}}_i$ is a vector spin operator of spin i, which can be explicitly written as $\{\hat{I}_{i,x}, \hat{I}_{i,y}, \hat{I}_{i,z}\}$.

The structure of the Hamiltonian in eq. (4) makes the role of intrapair permutation symmetry explicit. The first term, proportional to $J_{\text{gem}}$, as well as the second term proportional to $\Sigma J$, are symmetric under exchange of the two protons within each CH$_2$ group. Consequently, these terms commute with all operators that permute the two spins belonging to the same methylene unit. Their common eigenstates are therefore direct products of local singlet and triplet states of individual CH$_2$ groups:

$$\mathfrak{B}_{ST} = \left\{ \left|T_{+1}^{(12)}\right\rangle, \left|T_0^{(12)}\right\rangle, \left|S_0^{(12)}\right\rangle, \left|T_{-1}^{(12)}\right\rangle \right\} \otimes \left\{ \left|T_{+1}^{(34)}\right\rangle, \left|T_0^{(34)}\right\rangle, \left|S_0^{(34)}\right\rangle, \left|T_{-1}^{(34)}\right\rangle \right\} \otimes \ldots \tag{5}$$

Here the superscripts denote the spin numbers. When displaying product states such as $|S_0 T_0 T_0 T_0 \ldots\rangle$, we omit the spin indices, assuming that the order of states follows the numbering of spins.

By contrast, the third term in eq. (4), proportional to $\Delta J$, is antisymmetric under intrapair permutation. This term does not commute with the permutation operators and therefore couples product states that differ by the singlet–triplet character of one or more $CH_2$ units. As a result, the $\Delta J$ term introduces mixing between otherwise symmetry-adapted singlet–triplet product states. The nature of this mixing, and its consequences for the block structure of the Hamiltonian, are analyzed in detail below.

To make the coupling structure explicit, we first consider the minimal nontrivial system of two $CH_2$ groups ($n = 2$, four spins, $\dim(\widehat{H}_J) = 2^4 = 16$). We represent $\widehat{H}_J$ (eq. (4)) in the full singlet–triplet product basis $\mathcal{B}_{ST}$, i.e. the tensor product of $\{|T_{+1}\rangle, |T_0\rangle, |S_0\rangle, |T_{-1}\rangle\}$ states for each $CH_2$ group. In this basis, the Hamiltonian is almost entirely diagonal: all Zeeman contributions and the symmetric parts of the scalar couplings contribute only to diagonal terms. Remarkably, the matrix contains only two non-zero off-diagonal elements, both generated by the antisymmetric term proportional to $\Delta J$: a mixing between $|S_0 T_0\rangle$ and $|T_0 S_0\rangle$, and a mixing between $|S_0 S_0\rangle$ and $|T_0 T_0\rangle$.

These two mixings have qualitatively different roles. The $|S_0 T_0\rangle \leftrightarrow |T_0 S_0\rangle$ mixing connects states that are degenerate with respect to the dominant geminal energies (one singlet and one triplet unit) and therefore leads to strong hybridization already at zero order; we refer to this as *type-I mixing*. By contrast, the $|S_0 S_0\rangle \leftrightarrow |T_0 T_0\rangle$ mixing connects manifolds separated in energy by an amount set by $J_{\mathrm{gem}}$ (two singlets versus two triplets), so its principal effect is perturbative energy shifts; we refer to this as *type-II mixing*.

The same mixing pattern persists for longer chains. Adding additional $CH_2$ groups enlarges the basis by Kronecker products. As a result, the $\Delta J$ term systematically generates (i) nearest-neighbor couplings of the form $|\ldots S_0 T_0 \ldots\rangle \leftrightarrow |\ldots T_0 S_0 \ldots\rangle$ (type I), and (ii) couplings of the form $|\ldots S_0 S_0 \ldots\rangle \leftrightarrow |\ldots T_0 T_0 \ldots\rangle$ (type II), while all remaining product states remain uncoupled.

Since the only off-diagonal couplings involve $|S_0\rangle$ and $|T_0\rangle$, the zero-quantum dynamics of interest are confined to the subspace spanned by these states. The outer triplet states $|T_{+1}\rangle$ and $|T_{-1}\rangle$ remain isolated: they are split by Zeeman interactions and are not connected to the $\{|S_0\rangle, |T_0\rangle\}$ manifold. We therefore restrict each $CH_2$ group to an effective two-level system $\{|S_0\rangle, |T_0\rangle\}$, which enables the mapping to a chain of fictitious spin-½ particles (**Fig. 2d**).

At this stage it is convenient to make the energetic structure of the restricted Hilbert space explicit. In the absence of the inter-pair couplings proportional to $\Sigma J$ and $\Delta J$, the energies of the product states are determined solely by the intrapair geminal couplings. For a single $CH_2$ group, the eigenvalues of the operator $J_{\mathrm{gem}} \, \widehat{\mathbf{I}}_1 \cdot \widehat{\mathbf{I}}_2$ are $-\frac{3}{4} J_{\mathrm{gem}}$ for the singlet state $|S_0\rangle$ and $+\frac{1}{4} J_{\mathrm{gem}}$ for the triplet state $|T_0\rangle$.

For a product state containing $N_S$ singlet units and $N_T$ triplet units ($N_S + N_T = n$), the total contribution of the geminal couplings to the energy is therefore given by:

$$E_{gem} = J_{gem} \left( -\frac{3}{4} N_S + \frac{1}{4} N_T \right). \tag{6}$$

Here we express energy in the units of Hz. This expression makes explicit that manifolds differing by two singlet–triplet substitutions are separated in energy by an amount on the order of $|2 J_{gem}|$, justifying the perturbative treatment of the type-II mixing discussed above.

The analogy between the *J*-coupling Hamiltonian $\hat{H}_J$ (eq. (4)) and the XY Hamiltonian $\hat{H}_{XY}$ (eq. (1)) becomes explicit when the latter is expressed in the product basis:

$$\mathfrak{B}_{\alpha\beta} = \{|\alpha^{(1)}\rangle, |\beta^{(1)}\rangle\} \otimes \{|\alpha^{(2)}\rangle, |\beta^{(2)}\rangle\} \otimes \ldots \quad (7)$$

Here, superscripts denote the spin index, and are omitted in product states such as $|\beta\alpha\alpha\ldots\rangle$. In this basis, $\hat{H}_{XY}$ introduces a single type of mixing: neighboring $\alpha$ and $\beta$ states are exchanged, so that a state $|\ldots\alpha\beta\ldots\rangle$ is mixed with $|\ldots\beta\alpha\ldots\rangle$ through a matrix element $2\pi J/2$. This exchange mixing is directly analogous to the type-I mixing identified above for the Hamiltonian $\hat{H}_J$.

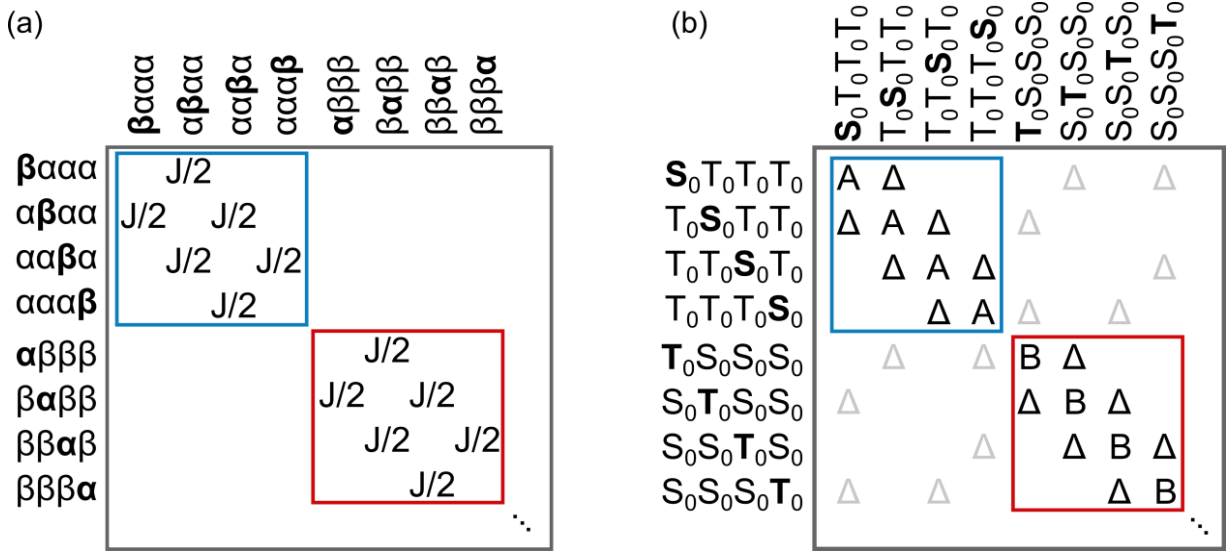

**Fig. 3. Block structure of the XY Hamiltonian and of the aliphatic-chain Hamiltonian.** Both matrices are divided by $2\pi$ and are shown in units of Hz. (a) Matrix representation of the $\hat{H}_{XY}$ Hamiltonian (eq. (1)) for the case of $n = 4$ spins. The ordering of product states in the $\mathfrak{B}_{\alpha\beta}$ basis is specified on the left-hand side and above the matrix. White spaces correspond to zero matrix elements. (b) Matrix representation of the $\hat{H}_J$ Hamiltonian (eq. (4)) for $n = 4$ CH$_2$ units, shown in the corresponding singlet–triplet product basis restricted to $\{|S_0\rangle, |T_0\rangle\}$ and ordered analogously. Type-I mixing is indicated by black off-diagonal elements equal to $\Delta J/2$, while type-II mixing is indicated by gray off-diagonal elements, also equal to $\Delta J/2$. Diagonal elements labeled A and B take the values $A = 0$ and $B = -2J_{\text{gem}}$, respectively, and occur repeatedly within the corresponding excitation manifolds.

As outlined in the Introduction, for the XY Hamiltonian $\hat{H}_{XY}$ (eq. (1)) the key conserved quantity is the total *z*-projection operator $\hat{I}_z$ (eq. (2)). This follows from the fact that $\hat{H}_{XY}$ commutes with $\hat{I}_z$, $[\hat{H}_{XY}, \hat{I}_z] = 0$. As a consequence, the Hamiltonian block-diagonalizes into disconnected subspaces, which we refer to as excitation subspaces or blocks. An excitation corresponds to a spin in the $|\beta\rangle$ state. For example, the single-excitation subspace contains basis states with one spin flipped, $\{|\beta\alpha\alpha\ldots\rangle, |\alpha\beta\alpha\ldots\rangle, \ldots\}$, while the three-excitation subspace contains states with three $|\beta\rangle$ spins, $\{|\beta\beta\beta\alpha\ldots\rangle, |\beta\beta\alpha\beta\ldots\rangle, \ldots\}$, and so on (see **Fig. 3a**).

A slightly more complicated structure is observed for the Hamiltonian $\hat{H}_J$ (eq. (4)). In addition to the type-I mixing, which is isomorphic to the mixing generated by $\hat{H}_{XY}$, the Hamiltonian $\hat{H}_J$ also contains type-II mixing. These off-diagonal terms connect excitation blocks whose numbers of $|S_0\rangle$ states differ by two (**Fig. 3b**). In the absence of type-II mixing, each block is identical to that of the XY Hamiltonian.

It is important to note, however, that the difference between diagonal matrix elements in neighboring excitation blocks is always $2J_{\text{gem}}$ (typical experimental values of $2J_{\text{gem}}$ fall in the range -26 to -30 Hz), which is much larger than the magnitude of the off-diagonal elements connecting these blocks ($\Delta J/2$ typically does not exceed 3–4 Hz; see, for example, ref. (*13*)). As a result, the inter-block couplings induced by type-II mixing can be treated perturbatively. To leading order, this mixing does not induce significant admixture between states belonging to different excitation manifolds, but it can lead to shifts of the eigenenergies.

In particular, situations arise in which two nearly degenerate states are coupled to the same outlying level. For example, the states $|T_0S_0S_0S_0\rangle$ and $|S_0S_0T_0S_0\rangle$ are both coupled via type-II mixing to the same outlying state $|T_0T_0T_0S_0\rangle$. This is a well-known scenario in quantum mechanics that is treated using degenerate perturbation theory, in which the degeneracy within the subspace is lifted at second order in the energies (*28*).

*Eigenstates and eigenenergies of tridiagonal Toeplitz matrices*

Consider a tridiagonal Toeplitz matrix shown in **Fig. 4a**. Such matrices have nonzero elements only on the principal diagonal and on the two diagonals immediately adjacent to it. When all elements along each of these diagonals are equal, the matrix is referred to as a Toeplitz matrix. The eigenstates of a real symmetric tridiagonal Toeplitz matrix are given by (*29*):

$$|\psi_k\rangle = \sqrt{\frac{2}{n+1}}\sum_{i=1}^{n} C(i,k,n)\,|i\rangle, \qquad (8)$$
$$C(i,k,n) = \sin\left(\frac{i\,k\,\pi}{n+1}\right).$$

Here, the basis states $|i\rangle$ correspond to the sorted product states of the excitation subspace, i.e. $\{|\beta\alpha\alpha\ldots\rangle, |\alpha\beta\alpha\alpha\ldots\rangle, \ldots\}$ for the XY model and $\{|S_0T_0T_0T_0\ldots\rangle, |T_0S_0T_0T_0\ldots\rangle, \ldots\}$ for the aliphatic chain. **Fig. 4b** shows the spatial distributions of the coefficients $C(i,k,n)$ for the first three eigenstates with wavenumbers $k = 1, 2, 3$ for the case of $n = 20$ spins.

Such delocalized eigenstates, formed as coherent superpositions of basis states extending over the entire spin chain, are rather unusual in liquid-state NMR. In practice, this delocalization manifests itself as a collective response of the system. We have previously demonstrated experimentally, for molecules containing three $CH_2$ groups, that application of a selective SLIC pulse to a terminal $CH_2$ unit leads to a simultaneous change of the spin state of all methylene groups in the chain, despite the fact that the $CH_2$ units are only weakly coupled (*12*).

The corresponding eigenenergies of a tridiagonal Toeplitz matrix are given by:

$$E_k = A + 2\Delta \cos\left(\frac{k\,\pi}{n+1}\right), \qquad (9)$$

where $A$ denotes the value of the principal diagonal elements, $\Delta$ is the value of the off-diagonal elements adjacent to the principal diagonal (see **Fig. 4a**), and $n$ is the dimension of the matrix. The index $k$ establishes a one-to-one correspondence between the eigenenergy $E_k$ and the delocalized eigenstate $|\psi_k\rangle$ given by eq. (8).

Eigenenergies calculated according to eq. (9) for spin chains containing from 2 to 20 nodes are shown in **Fig. 4c**, with $A = 0$. The total number of eigenstates equals the number of nodes $n$. For $n = 2$, two levels separated by $2\Delta$ are obtained, corresponding to $J$ for the XY model and to $\Delta J$ for the aliphatic chain; for $n = 3$, three levels separated by $2\Delta/\sqrt{2}$ appear, corresponding to $J/\sqrt{2}$ and $\Delta J/\sqrt{2}$, respectively. In the limit of large $n$, the spectral bandwidth approaches $4\Delta$. As discussed below, transitions between arbitrary pairs of levels can, in principle, be excited and observed.

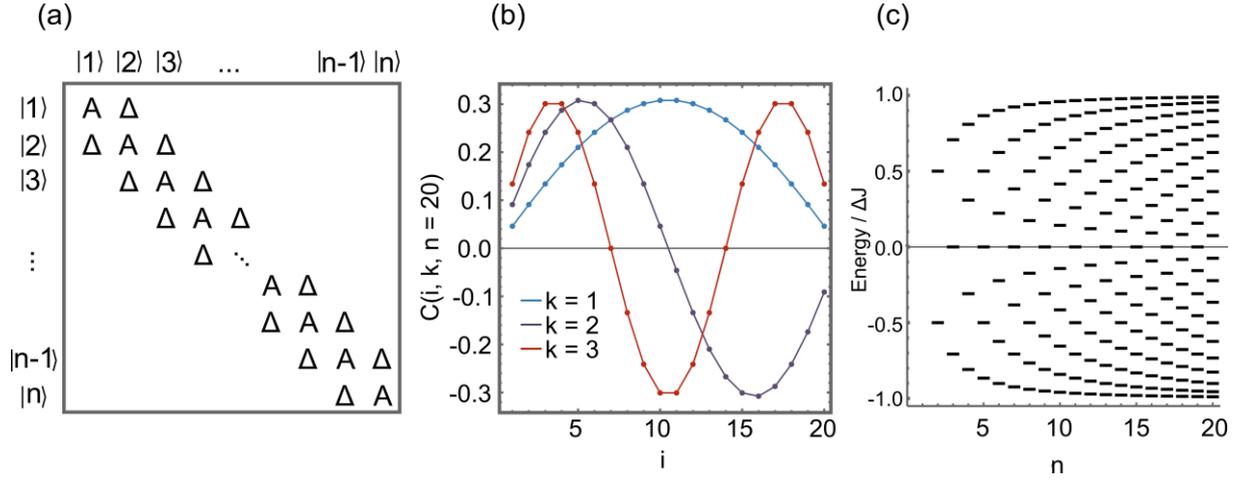

**Fig. 4**. **Eigenstates and eigenenergies of tridiagonal Toeplitz matrices**. (a) General form of the tridiagonal Toeplitz matrices that appear as blocks in the single-excitation subspaces of the $\hat{H}_{XY}$ and $\hat{H}_J$ Hamiltonians (see text for explanation). (b) Visualization of the coefficients of the eigenstates distributed over the basis states $|1\rangle, |2\rangle, ..., |n\rangle$ for the case of $n = 20$ spins. The eigenstate coefficients depend neither on $A$ nor on $\Delta$. The eigenstates are characterized by the wavenumber $k$ and form patterns reminiscent of standing waves. (c) Eigenenergies of the tridiagonal Toeplitz matrices calculated according to eq. (9) for chain lengths $n$ ranging from 2 to 20. The diagonal elements were set to $A = 0$.

*Higher-excitation subspaces: two-excitation block*

Spin chains with $n \geq 4$ nodes exhibit Hamiltonian blocks that are more complex than those shown in **Fig. 3**. An example of such excitation blocks is shown in **Fig. 5** for the case of $n = 4$. Specifically, **Fig. 5a** displays the two-excitation subspace of the XY model, while **Fig. 5b** shows the corresponding block structure of the aliphatic-chain Hamiltonian, which involves mixing between the two-, four-, and zero-excitation subspaces. To the best of our knowledge, no general analytical expressions for the eigenstates and eigenenergies of such finite-dimensional matrices are available.

Even though Feynman notes that in the case of an infinite spin chain the problem of two spin waves admits an exact analytical solution:

"Interestingly enough, an exact solution can be written down if there are just the *two* down spins. ... This solution includes the "interaction" of the two spins. It describes the fact that when the spins come together there is a certain chance of scattering. The spins act very much like particles with an interaction." (6).

This solution does not directly apply to finite spin chains in which the first and last spins are not coupled to each other and whose Hamiltonians have a block structure of the type considered above. As a consequence, one may expect additional transition frequencies to appear in spin-chain zero-quantum NMR spectra when higher-excitation subspaces become populated.

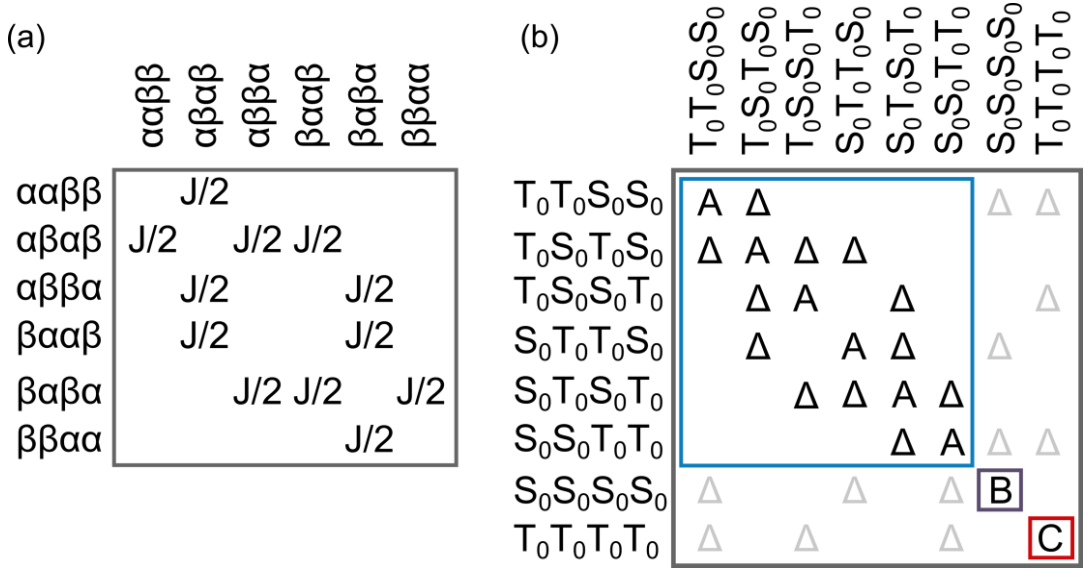

**Fig. 5. Higher-excitation Hamiltonian blocks for a four-site spin chain.** (a) Matrix representation of the two-excitation subspace of the XY Hamiltonian $\hat{H}_{XY}$ (eq. (1)) for $n = 4$ spins. The ordering of the product states in the $\mathfrak{B}_{\alpha\beta}$ basis is indicated on the left-hand side and above the matrix. (b) Corresponding matrix representation of the aliphatic-chain Hamiltonian $\hat{H}_J$ (eq. (4)) for $n = 4$ CH$_2$ units, shown in the singlet–triplet product basis restricted to $\{|S_0\rangle, |T_0\rangle\}$. In contrast to the XY case, the two-excitation subspace is coupled to the zero- and four-excitation subspaces via type-II mixing, leading to a more complex block structure. A, B, and C label distinct diagonal matrix elements: A = -$J_{gem}$, B = -3$J_{gem}$, C = $J_{gem}$; Δ = ΔJ/2.

## Results

In order to test the theoretical predictions, we performed numerical simulations of spin dynamics in XY spin chains and in spin chains formed by aliphatic molecules. Time-domain signals were calculated by solving the Liouville–von Neumann equation for the evolution of the density operator (*30*). Initial density operator for the XY model was set in the form of:

$$\hat{\rho}(0) = -\hat{I}_{1,z} + \hat{I}_{2,z} + \hat{I}_{3,z} + \hat{I}_{4,z}, \tag{10}$$

Here we show an example for the case $n = 4$, in which the first spin is inverted while the remaining spins are polarized along the *z*-axis. Simulations were performed for different chain lengths, from $n = 2$ to $n = 5$, and for different configurations of initial spin inversions, as shown in the Supplementary Materials.

Initial density operator for the aliphatic spin chain was set in the form of:

$$\hat{\rho}(0) = |T_0S_0S_0S_0\rangle\langle T_0S_0S_0S_0| + |S_0T_0S_0S_0\rangle\langle S_0T_0S_0S_0| + \\ |S_0S_0T_0S_0\rangle\langle S_0S_0T_0S_0| - |S_0S_0S_0T_0\rangle\langle S_0S_0S_0T_0|, \tag{11}$$

Here we again consider an example of a system with $n = 4$ CH$_2$ groups, corresponding to three fictitious spins polarized in one direction and one terminal spin inverted in the opposite direction.

The evolution of the density operator was calculated using the well-known solution of the Liouville–von Neumann equation for time-independent Hamiltonians:

$$\hat{\rho}(t) = exp(-i\hat{H}t)\hat{\rho}(0)exp(i\hat{H}t), \tag{12}$$

For the XY model, the Hamiltonian was set to $\hat{H}_{XY}$ (eq. (1)), while for the aliphatic chain it was set to $\hat{H}_J$ (eq. (4)). The time-domain signal was obtained by calculating the expectation value of an observation operator $\hat{O}$:

$$\langle \hat{O}(t) \rangle = Tr\left(\hat{O}\hat{\rho}(t)\right), \tag{13}$$

For the XY model, the observation operator $\hat{O}$ was chosen as $\hat{I}_{i,z}$, where $i \in \{1,2,...,n\}$. For the aliphatic chain, the observation operator was taken as the population of an individual product state; for example, for $n = 4$, this corresponds to the population of the state $|T_0 S_0 S_0 S_0\rangle\langle T_0 S_0 S_0 S_0|$ for $i = 1$, or $|S_0 T_0 S_0 S_0\rangle\langle S_0 T_0 S_0 S_0|$ for $i = 2$, and so on.

In all simulations, the *J*-coupling in the XY chain was set to 5 Hz. For the aliphatic chain, the value of $\Delta J$ was set to 5 Hz, while $J_{\text{gem}}$ was set to -14 Hz. Simulations were performed using two programs: for spin systems containing fewer than six spins, calculations were carried out using SpinDynamica (*31*), whereas for spin systems containing six to eight spins, calculations were performed using home-written Wolfram Mathematica code. As a result, we obtained time-domain trajectories similar to those shown in **Fig. 1**; the only difference is that the signals were calculated over a time interval of up to 20 s. The signals were sampled with a time step of 0.005 s.

Phenomenological relaxation was introduced by multiplying the time-domain signal by a monoexponentially decaying function with a relaxation time constant $\tau = 5$ s. Fourier transformation was then performed after subtracting the DC component of the signal in order to avoid a strong peak at zero frequency. The resulting spin-chain spectra were displayed in magnitude mode. The results of numerical simulations were compared with the theoretical predictions for the transition frequencies, calculated as follows:

$$\nu_{kl} = E_k - E_l, \tag{14}$$

where the energies $E_k$ and $E_l$ were calculated according to eq. (9).

**Fig. 6** shows the results of numerical simulations for the case of $n = 4$ spins in the XY model and $n = 4$ $CH_2$ units in the aliphatic spin chain. Red dashed lines denote the expected transition frequencies calculated according to eq. (14). For the XY model, four transitions are expected and are indeed observed in the numerical simulations (**Fig. 6c**). In this panel, the initial density operator $\hat{\rho}(0)$ was set according to eq. (10), and the observation operator $\hat{O}$ was chosen as $\hat{I}_{1,z}$, such that the first spin was initially flipped and also observed.

For the aliphatic spin chain, the initial density operator was set according to eq. (11), and the observation operator $\hat{O}$ was chosen as the population operator $|S_0 S_0 S_0 T_0\rangle\langle S_0 S_0 S_0 T_0|$, so that the observation corresponds to the initially flipped fictitious spin-½. As can be seen, the aliphatic spin-chain spectrum exhibits two transitions that are split into two spectral components (**Fig. 6a**). Specifically, the transitions $\nu_{12}$ and $\nu_{34}$, which are degenerate in the XY model, become non-degenerate in the aliphatic chain due to the perturbation introduced by the $\hat{H}_J$ Hamiltonian, as illustrated in **Fig. 3b**. The same behaviour is observed for the transitions $\nu_{13}$ and $\nu_{24}$, which also split. In contrast, the transitions $\nu_{23}$ and $\nu_{14}$ coincide for the XY model and the aliphatic spin chain. This behavior can be understood by noting that the outer energy levels labeled 1 and 4 shift relative to the central levels 2 and 3 (see **Fig. 6e**).

**Fig. 6b** and **Fig. 6d** show the results for the case in which the initial density operator was set to $-\hat{I}_{1,z} - \hat{I}_{2,z} + \hat{I}_{3,z} + \hat{I}_{4,z}$ for the XY model and to $|S_0 S_0 T_0 S_0\rangle\langle S_0 S_0 T_0 S_0| + |S_0 S_0 S_0 T_0\rangle\langle S_0 S_0 S_0 T_0|$ for the aliphatic chain. The observation was again performed on the terminal node. In this case, the intensities of individual transitions differ from those in **Fig. 6a** and **Fig. 6c**, with the most pronounced change being that the transitions $\nu_{13}$ and $\nu_{24}$ are essentially suppressed in both cases. No additional spectral components are observed.

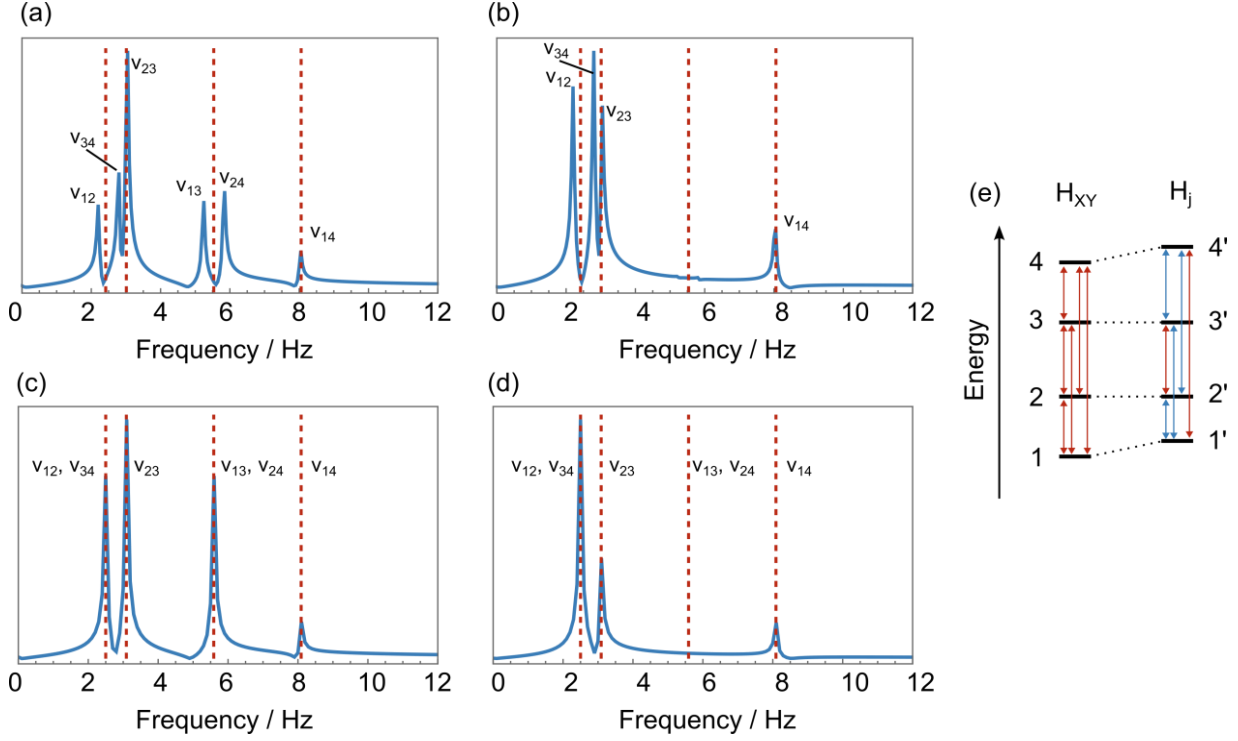

**Fig. 6**. **Spin-chain spectra for the case of n = 4 spins (XY model) and n = 4 CH$_2$ units (aliphatic chain), with the observation operator $\hat{O}$ applied to the terminal site**. (a) Numerical simulation of the spin-chain zero-quantum NMR spectrum for an aliphatic chain with the initial density operator given by eq. (11). Red dashed vertical lines indicate the theoretically predicted transition frequencies calculated according to eqs. (9) and (14). (b) Same as panel (a), except that the initial density operator $\hat{\rho}(0)$ was set to $|S_0S_0T_0S_0\rangle\langle S_0S_0T_0S_0| + |S_0S_0S_0T_0\rangle\langle S_0S_0S_0T_0|$. (c) Same as panel (a), except that the spin system corresponds to the XY model. The evolution occurs under $\hat{H}_{XY}$ (eq. (1)), in contrast to $\hat{H}_J$ (eq. (4)) used for the panel (a). The initial density operator $\hat{\rho}(0)$ was set according to eq. (10), corresponding to the case in which the first spin is initially inverted and the evolution of its z-component is observed. (d) Same as panel (c), except that the initial density operator was set to $\hat{\rho}(0) = -\hat{I}_{1,z} - \hat{I}_{2,z} + \hat{I}_{3,z} + \hat{I}_{4,z}$. (e) Energy-level diagrams for the XY model and the aliphatic chain, illustrating the shift of the outer levels relative to the central ones. Six possible transitions in the XY model are indicated by red arrows. Some transitions are degenerate due to the symmetry properties of the eigenstates given by Eq. (9). This degeneracy is lifted in the aliphatic chain. Blue arrows indicate transition frequencies that differ between the XY model and the aliphatic chain, while red arrows indicate transitions that occur at the same frequencies in both models. The shift of the energy levels in the aliphatic chain arises from the perturbation induced by type-II mixing in the $\hat{H}_J$ Hamiltonian, as illustrated in **Fig. 3b**. Simulation parameters were as follows: $J = 5$ Hz, $\Delta J = 5$ Hz, $J_{gem} = -14$ Hz, and relaxation time constant $\tau = 5$ s. See the main text for additional details of the numerical simulations.

**Fig. 7** shows the results of numerical simulations of XY spin-chain spectra for different chain lengths ranging from $n = 2$ to $n = 5$. In all cases, the initial density operator $\hat{\rho}(0)$ was chosen such that the first spin is inverted with respect to the remaining spins. For each value of $n$, the observation operator $\hat{O}$ was taken as $\hat{O}_i = \hat{I}_{i,z}$, with $i \in \{1, 2, \dots, n\}$.

In agreement with the theoretical predictions, the frequencies of the observed spectral components coincide with those calculated according to eq. (14). In some cases, not all frequency modes are excited; for example, for $n = 3$ and $i = 2$, only one spectral component out of the two possible transitions appears in the spectrum, corresponding to $\nu_{13}$.

Additionally, in all cases a symmetry is observed between the spectra detected at the $i$-th node and at the $(n + 1 - i)$-th node. This behavior reflects the mirror symmetry of the spin chain with respect to its central site. While the corresponding spin evolutions may differ in phase, such phase differences are not visible when the spectra are displayed in magnitude mode.

Taken together, these results demonstrate good agreement between the theoretical framework developed above and the numerical simulations, providing a basis for understanding the spectroscopy of spin-wave–like excitations that can be observed in molecules containing aliphatic chains with an arbitrary number of weakly coupled $CH_2$ units.

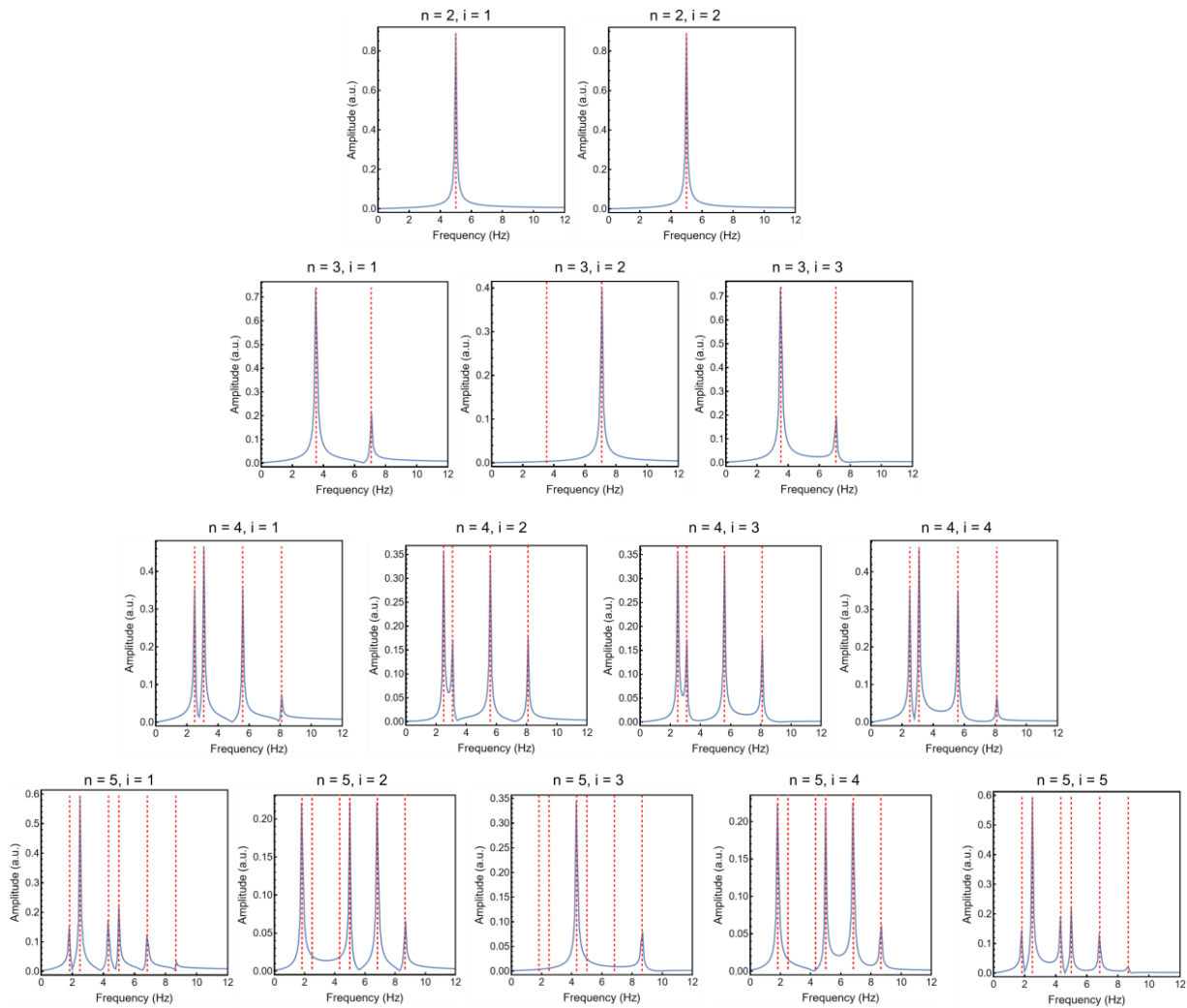

**Fig. 7. Numerical simulations of XY spin-chain spectra for chain lengths ranging from n = 2 to n = 5 spins.** In all cases, the initial density operator $\hat{\rho}(0)$ was chosen such that the first spin is inverted with respect to the remaining spins: $-\hat{I}_{1,z} + \hat{I}_{2,z}$ ($n = 2$), $-\hat{I}_{1,z} + \hat{I}_{2,z} + \hat{I}_{3,z}$ ($n = 3$), $-\hat{I}_{1,z} + \hat{I}_{2,z} + \hat{I}_{3,z} + \hat{I}_{4,z}$ ($n = 4$), and $-\hat{I}_{1,z} + \hat{I}_{2,z} + \hat{I}_{3,z} + \hat{I}_{4,z} + \hat{I}_{5,z}$ ($n = 5$). The observation operator $\hat{O}$ was varied from the first to the $n$-th spin, which is reflected by the index $i$ indicated for each panel. For example, the top row corresponds to the case $n = 2$, with observation performed on the first spin ($i = 1$) and on the second spin ($i = 2$). All spectra are displayed in magnitude mode. Simulation parameters were $J = 5$ Hz and relaxation time constant $\tau = 5$ s. See the main text for additional details of the numerical simulations.

## Discussion

The theory presented here can explain previously reported experimental results (*21, 22*). These observations are summarized in **Fig. 8**. In these experiments, a non-uniform distribution of singlet-state populations was created along the aliphatic chain of the DSS molecule (2,2-dimethyl-2-silapentane-5-sulfonate sodium salt), which is commonly used as an NMR standard for chemical-shift refence in aqueous solutions. This molecule contains three well-separated $CH_2$ groups, and each pair of geminal protons is magnetically inequivalent, thus fulfilling the theoretical conditions specified in the Theory section.

In the experiment shown in **Fig. 8a**, overpopulation of singlet states was created on CH$_2$ groups $i = 1$ and 2, after which decoupling of the first CH$_2$ group was applied during the evolution period of collective zero-quantum coherences. This decoupling effectively reduced the spin system from $n = 3$ to $n = 2$. In this case, a single strong spectral component was observed at a frequency corresponding to $\Delta J$ between methylene groups $i = 2$ and 3. This observation is in agreement with the $n = 2$ case illustrated in the top row of **Fig. 7**.

Another experiment on DSS is shown in **Fig. 8b**. In this case, all experimental parameters were identical, except that no decoupling was applied during the zero-quantum evolution period. Three spectral components were observed, in good agreement with the case shown in the second row of **Fig. 7**. The main difference is that the low-frequency spectral component appearing at approximately 4 Hz is split into two components with frequencies $\nu_{12} = 4.67$ Hz and $\nu_{23} = 3.70$ Hz, while the higher-frequency component at $\nu_{13} = 8.37$ Hz matches the sum $\nu_{12} + \nu_{23}$. This behavior is expected, as the evolution occurs under the Hamiltonian $\hat{H}_J$, which includes type-II mixing that leads to splitting of the $\nu_{12}$ and $\nu_{23}$ transitions. Similar behavior is observed in **Fig. 6**.

Finally, it should be noted that the DSS molecule does not strictly correspond to an idealized spin chain, since both geminal and vicinal *J*-couplings vary slightly along the chain (experimentally measured couplings are reported in Refs. (*12*, *22*)). As a result, the splitting between the $\nu_{12}$ and $\nu_{23}$ lines deviates slightly from the second-order degenerate-perturbation-theory estimate $\frac{1}{4}(\Delta J)^2/J_{\text{gem}}$ (*22*). Experimental details for the data shown in **Fig. 8** are summarized in the Methods section and are essentially identical to those reported previously in Ref. (*22*).

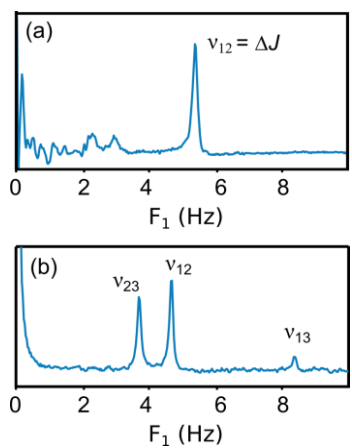

**Fig. 8**. **Experimental spin-chain zero-quantum NMR spectra**. (a) Spectrum acquired for DSS (2,2-dimethyl-2-silapentane-5-sulfonate sodium salt) with decoupling of one of the CH$_2$ groups applied during the evolution of zero-quantum spin-chain coherences. In this case, the effective spin system corresponds to $n = 2$ CH$_2$ groups. In agreement with the theoretical treatment developed in this work, a single intense spectral component is observed at the frequency corresponding to $\Delta J$. Two weak low-frequency features are also present, which are likely associated with decoupling-related artifacts. (b) Spectrum acquired for DSS without decoupling during the free evolution period. In this case, the spin system corresponds to $n = 3$ CH$_2$ groups, and three spectral components are observed.

An important question is which real molecular systems can be reasonably described by the idealized aliphatic spin-chain model introduced here. To address this question, we outline several classes of molecular systems that appear experimentally feasible.

The first category consists of achiral molecules of the general form R$_1$–(CH$_2$)$_n$–R$_2$. The substituent R$_1$ should be a σ-electron–donating group, such as an alkyl group (–CH$_3$), silyl (–SiR$_3$), stannyl or germyl groups, or thioalkyl groups (–SR). At the same time, R$_2$ should be a σ-electron–withdrawing group, such as perfluoroalkyl groups (including –CF$_3$ and (CF$_2$)$_n$–CF$_3$). Because fluorine is highly electronegative, multiple C–F bonds strongly withdraw σ-electron density, leading to substantial chemical-shift dispersion. This provides an interesting scenario in which a spin chain may be constructed from both protonated and fluorinated methylene groups. Indeed, it has recently been shown that singlet states of fluorine nuclei can be created in a manner analogous to those of protonated methylene groups (*32*).

Other possible substituents for $R_2$ include sulfonyl-based groups ($-SO_2R$, $-SO_3H$, $-SO_3^-$), carbonyl-derived groups such as esters ($-CO_2R$), carboxylic acids ($-CO_2H$), amides ($-CONR_2$), ketones and aldehydes ($-COR$), as well as nitriles ($-CN$), halogens ($-F$, $-Cl$, $-Br$, $-I$), nitro groups ($-NO_2$), positively charged groups ($-NR_3^+$, $-PR_4^+$, $-SR_2^+$), and ether or thioether groups ($-OR$, $-SR$). For example, DSS, which has $R_1 = -SO_3^-$ and $R_2 = -Si(CH_3)_3$, exhibits chemical shifts spread over approximately 2 ppm. Another example is lysine, which contains four well-resolved $CH_2$ groups with chemical shifts spanning roughly 1.5–3 ppm. Lysine is a chiral molecule, which makes the theoretical treatment more involved, but it nevertheless illustrates the feasibility of resolving multiple methylene units.

It should be possible to synthesize molecules with up to six well-resolved methylene groups, particularly at high magnetic fields (e.g., 1.2 GHz, corresponding to the highest-field NMR spectrometers currently available). With respect to scalar couplings, such systems are expected to exhibit relatively uniform *J*-couplings along the chain, since the local structure of each $CH_2$ unit is identical.

The second category includes fluorinated molecules of the form $R_1-(CF_2)_n-R_2$. We have recently found that perfluorinated methylene groups possess accessible singlet states, and that the dispersion of chemical shifts is typically larger for fluorine nuclei (*32*). A third category comprises mixed systems of the general form $R_1-(CH_2)_n-(CF_2)_m-R_2$. At the highest available magnetic fields, one may expect to resolve up to six protonated methylene groups and up to twelve methylene units in mixed fluorinated–protonated chains.

Finally, alternative systems include molecules containing paramagnetic tags attached to one end of an aliphatic chain. It has been shown that paramagnetic pseudocontact-shift agents, such as Co(II) complexes attached to one terminus of a molecule, can generate large chemical-shift dispersions, allowing up to 15 (Ref. (*33*)) or even 16 (Ref. (*34*)) resolved $CH_2$ groups. These systems therefore provide realistic molecular platforms in which the effective spin-½ description of $CH_2$ units and their mapping onto one-dimensional XY spin chains can be explored experimentally.

We now consider several potential applications of spin-chain zero-quantum NMR experiments. A first application concerns the extraction of information about conformational dynamics associated with rotations of aliphatic chains about C–C bonds. In particular, the appearance of spin-chain zero-quantum NMR spectra would allow one to determine how many $CH_2$ units are effectively coupled through non-zero $\Delta J$ couplings within the spin chain. As shown above, spin-chain spectra exhibit different numbers of spectral components depending on the number *n* of methylene units involved (see **Fig. 7**). These spectra therefore provide a characteristic fingerprint of the number of $CH_2$ groups that are coupled through non-zero $\Delta J$ interactions. Such information is challenging to obtain using conventional NMR approaches.

Furthermore, $\Delta J$ couplings provide direct insight into conformational rotations of $CH_2$ groups about C–C bonds (*35*, *36*). It should be possible to develop computational approaches to extract $\Delta J$ couplings by combining the analysis of conventional one-dimensional NMR spectra—which are often crowded and broadened by *J*-coupling splittings—with the analysis of comparatively sparse spin-chain zero-quantum NMR spectra. The latter benefit from longer relaxation time constants and complete insensitivity to static magnetic-field inhomogeneities (*21*, *22*).

A second class of applications lies in the domain of quantum physics, specifically in the experimental investigation of delocalization phenomena associated with wave functions that extend over multiple particles. The eigenstates described by eq. (8) belong to the class of non-separable, or entangled, multi-spin states (*37*, *38*), which are of central interest in quantum information theory. Finally, the experimental realization and observation of collective coherent dynamics of singlet–triplet spin waves may provide a useful testbed for benchmarking coherent many-body dynamics relevant to quantum information processing (*24*).

We briefly discuss spin-chain zero-quantum NMR in the context of related experimental observations. As shown in this work, zero-quantum coherences in aliphatic spin chains can be understood as well-defined collective spin-wave modes. Experimentally, it has been observed for the cases $n = 2$ and $n = 3$ that these coherences exhibit unusual relaxation properties (*21*, *22*). Specifically, their lifetimes can exceed both the transverse ($T_2$) and even the longitudinal ($T_1$) relaxation times of the associated $CH_2$ groups, thereby forming so-called collective long-lived coherences (*22*). Historically, long-lived coherences were first discovered in two-spin systems (*39–41*), but have since been extended to multi-spin systems. Investigating the lifetimes and relaxation mechanisms of such collective coherences is therefore of fundamental interest.

Closely related spin physics arises in zero- to ultralow-field NMR *J*-spectroscopy (*42*), where conservation of total spin in heteronuclear systems leads to coherent oscillations of individual spin polarizations when the nuclei begin with unequal initial polarization. Another system exhibiting related spin dynamics is signal amplification by reversible exchange (SABRE) at zero magnetic fields (*43–45*). In these experiments, initial singlet spin order originating from parahydrogen is transferred to spin polarization of a substrate molecule within a transient iridium-based complex. Together, these examples highlight the broader relevance of collective singlet–triplet dynamics across a range of NMR methodologies.

## Perspectives

It can be shown that, in the case of strongly coupled $CH_2$ groups, strong coupling does not suppress spin-wave–like behavior but instead leads to more complex spin dynamics, since the $|T_{+1}\rangle$ and $|T_{-1}\rangle$ states of individual $CH_2$ units can no longer be neglected. In this regime, the theory of spin chains should be extended to describe coupled four-level units rather than effective two-level systems.

Additional extensions of the theoretical framework can also be developed to account for complications arising from the presence of chiral centers in molecules. In this context, particularly interesting systems include amino acids such as glutamine and glutamate ($n = 2$), proline ($n = 3$), and lysine ($n = 4$). Modifications of spin-chain zero-quantum NMR experiments applied to methylene groups in these molecules are of considerable importance for probing the conformational dynamics of amino-acid side chains, especially in the field of intrinsically disordered proteins (IDPs). Indeed, even studies of the lifetimes of singlet states in glycine residues ($n = 1$) have recently led to significant advances in understanding the dynamics of IDPs (*46*).

## Methods

NMR experiments were performed at 298 K using a 5-mm iProbe on a Bruker 500 MHz wide-bore spectrometer ($B_0$ = 11.7 T) equipped with a Neo console. The sample contained approximately 50 mM DSS dissolved in $D_2O$ and was measured without degassing.

Spin-chain zero-quantum NMR spectra shown in **Fig. 8** were obtained using a polychromatic SLIC pulse sequence (*12*, *22*). The sequence began with a non-selective 90° pulse, followed by a polychromatic SLIC irradiation that addressed $CH_2$ groups 1 and 2, while $CH_2$ group 3 was simultaneously decoupled. After excitation, a variable evolution delay was introduced to allow the collective spin-wave coherences to evolve. For the experiment shown in **Fig. 8a**, an additional decoupling RF field was applied during this period, effectively reducing the spin system from six to four spins. In contrast, no pulses were applied during the evolution period for the experiment shown in **Fig. 8b**.

Following the evolution delay, a $T_{00}$ filter (*47*) was applied to suppress unwanted coherences. The filter consisted of three Gz gradient pulses, each followed by a 200 ms delay and interleaved with three non-selective 90° radiofrequency pulses. The gradient pulses had sinusoidal shapes with durations of 4.4, 2.4, and 2.0 ms and amplitudes of 10, −10, and -15% of the maximum available gradient strength (Gz = 0.5 T/m), respectively. The first RF pulse of the $T_{00}$ filter, applied between the first and second gradients, had a phase ϕ = 90° (along the y-axis), while the second and third pulses were applied after the second and third gradients with phases ϕ = 54.7° and 0°, respectively.

After coherence filtering, a reconversion SLIC pulse was applied selectively to $CH_2$ group 1 for signal detection. The SLIC pulses used for reconversion had an amplitude $ν_{SLIC}$ = 27 Hz and a duration $τ_{SLIC}$ = 110 ms, whereas excitation SLIC pulses were applied with $ν_{SLIC}$ = 14 Hz and $τ_{SLIC}$ = 155 ms. Nutation frequency for the decoupling was always set to 100 Hz. Polychromatic SLIC irradiation was implemented through superposition of rectangular, phase-modulated pulses.

All experiments employed a four-step phase cycle, in which the phases of the excitation SLIC pulses were alternated as (y, −y, y, −y) and those of the reconversion SLIC pulses as (y, y, −y, −y), while the receiver phase followed the pattern (y, −y, −y, y)(*48*). Recovery delays were set to five times the longitudinal relaxation time ($5T_1 = 10$ s). Zero-quantum spectra were acquired using four transients, with a maximum evolution time $t_{1,\text{max}} = 17$ s and 1024 increments at a step size $\Delta t_1 = 16.7$ ms. The resulting $t_1$-dependent oscillatory signals arising from collective long-lived coherences were converted to the $\omega_1$ frequency domain using a real (cosine) Fourier transformation.

## Conclusions

We have developed a generalized theoretical framework that reveals a previously unexplored class of spin-wave–like modes, corresponding to the coherent and magnetically silent propagation of singlet–triplet population imbalances along aliphatic chains. This theoretical treatment enables the determination of eigenstates and eigenfrequencies of zero-quantum collective long-lived coherences in chains containing an arbitrary number of chemically resolved $CH_2$ units, thereby forming the basis for a novel type of spin-chain zero-quantum NMR spectroscopy.

We have demonstrated that the proposed theoretical approach successfully explains previously reported experimental observations for systems containing $n = 2$ and $n = 3$ $CH_2$ groups. In addition, we have discussed experimentally feasible molecular systems in which analogous dynamics may be explored in significantly longer aliphatic chains. Finally, a range of potential applications and future perspectives has been outlined, spanning NMR-based studies of aliphatic-chain structure and conformational dynamics, as well as collective quantum spin phenomena.

## Acknowledgements

The author thanks John W. Blanchard for providing Wolfram Mathematica code developed at Berkeley for calculations in zero- and ultralow-field NMR. The code was adapted in this work for simulations of spin-chain zero-quantum NMR. The author also acknowledges Dmitry A. Cheshkov for helpful discussions regarding suitable molecular substituents for aliphatic chains that maximize chemical-shift dispersion.

The original draft of the manuscript was written by the author. Language and stylistic editing were refined through an iterative process using ChatGPT as a language model. Assistance from ChatGPT was also used in the development of selected SpinDynamica and Wolfram Mathematica routines, which are labeled accordingly in the code; all generated routines were subsequently verified by the author.

## Financial support

We are grateful for the support from l'Agence Nationale de la Recherche (ANR) on the project THROUGH-NMR (grant no. ANR-24-CE93-0011-01). We are indebted to the CNRS for support in terms of salary, and to the ENS for generous laboratory space.

## Supplementary Materials

All computer programs developed in this work are available as Supplementary Material at DOI: https://doi.org/10.5281/zenodo.18076308

Yosri, G. Young, A. Zalcman, C. Zhang, Y. Zhang, N. Zhu, N. Zobrist, Google Quantum AI and Collaborators, Observation of constructive interference at the edge of quantum ergodicity. *Nature* **646**, 825–830 (2025).

24. C. Zhang, R. G. Cortiñas, A. H. Karamlou, N. Noll, J. Provazza, J. Bausch, S. Shirobokov, A. White, M. Claassen, S. H. Kang, A. W. Senior, N. Tomašev, J. Gross, K. Lee, T. Schuster, W. J. Huggins, H. Celik, A. Greene, B. Kozlovskii, F. J. H. Heras, A. Bengtsson, A. G. Dau, I. Drozdov, B. Ying, W. Livingstone, V. Sivak, N. Yosri, C. Quintana, D. Abanin, A. Abbas, R. Acharya, L. A. Beni, G. Aigeldinger, R. Alcaraz, S. Alcaraz, T. I. Andersen, M. Ansmann, F. Arute, K. Arya, W. Askew, N. Astrakhantsev, J. Atalaya, B. Ballard, J. C. Bardin, H. Bates, M. B. Karimi, A. Bilmes, S. Bilodeau, F. Borjans, A. Bourassa, J. Bovaird, D. Bowers, L. Brill, P. Brooks, M. Broughton, D. A. Browne, B. Buchea, B. B. Buckley, T. Burger, B. Burkett, J. Busnaina, N. Bushnell, A. Cabrera, J. Campero, H.-S. Chang, S. Chen, Z. Chen, B. Chiaro, L.-Y. Chih, A. Y. Cleland, B. Cochrane, M. Cockrell, J. Cogan, R. Collins, P. Conner, H. Cook, W. Courtney, A. L. Crook, B. Curtin, S. Das, M. Damyanov, D. M. Debroy, L. D. Lorenzo, S. Demura, L. B. D. Rose, A. D. Paolo, P. Donohoe, A. Dunsworth, V. Ehimhen, A. Eickbusch, A. M. Elbag, L. Ella, M. Elzouka, D. Enriquez, C. Erickson, V. S. Ferreira, M. Flores, L. F. Burgos, E. Forati, J. Ford, A. G. Fowler, B. Foxen, M. Fukami, A. W. L. Fung, L. Fuste, S. Ganjam, G. Garcia, C. Garrick, R. Gasca, H. Gehring, R. Geiger, É. Genois, W. Giang, C. Gidney, D. Gilboa, J. E. Goeders, E. C. Gonzales, R. Gosula, S. J. de Graaf, D. Graumann, J. Grebel, J. Guerrero, J. D. Guimarães, T. Ha, S. Habegger, T. Hadick, A. Hadjikhani, M. P. Harrigan, S. D. Harrington, J. Hartshorn, S. Heslin, P. Heu, O. Higgott, R. Hiltermann, J. Hilton, H.-Y. Huang, M. Hucka, C. Hudspeth, A. Huff, E. Jeffrey, S. Jevons, Z. Jiang, X. Jin, C. Joshi, P. Juhas, A. Kabel, H. Kang, K. Kang, R. Kaufman, K. Kechedzhi, T. Khattar, M. Khezri, S. Kim, R. King, O. Kiss, P. V. Klimov, C. M. Knaut, B. Kobrin, F. Kostritsa, J. M. Kreikebaum, R. Kudo, B. Kueffler, A. Kumar, V. D. Kurilovich, V. Kutsko, N. Lacroix, D. Landhuis, T. Lange-Dei, B. W. Langley, P. Laptev, K.-M. Lau, L. L. Guevel, J. Ledford, J. Lee, B. J. Lester, W. Leung, L. Li, W. Y. Li, M. Li, A. T. Lill, M. T. Lloyd, A. Locharla, D. Lundahl, A. Lunt, S. Madhuk, A. Maiti, A. Maloney, S. Mandra, L. S. Martin, O. Martin, E. Mascot, P. M. Das, D. Maslov, M. Mathews, C. Maxfield, J. R. McClean, M. McEwen, S. Meeks, K. C. Miao, R. Molavi, S. Molina, S. Montazeri, C. Neill, M. Newman, A. Nguyen, M. Nguyen, C.-H. Ni, M. Y. Niu, L. Oas, R. Orosco, K. Ottosson, A. Pagano, S. Peek, D. Peterson, A. Pizzuto, E. Portoles, R. Potter, O. Pritchard, M. Qian, A. Ranadive, M. J. Reagor, R. Resnick, D. M. Rhodes, D. Riley, G. Roberts, R. Rodriguez, E. Ropes, E. Rosenberg, E. Rosenfeld, D. Rosenstock, E. Rossi, D. A. Rower, M. S. Rudolph, R. Salazar, K. Sankaragomathi, M. C. Sarihan, K. J. Satzinger, M. Schaefer, S. Schroeder, H. F. Schurkus, A. Shahingohar, M. J. Shearn, A. Shorter, N. Shutty, V. Shvarts, S. Small, W. C. Smith, D. A. Sobel, R. D. Somma, B. Spells, S. Springer, G. Sterling, J. Suchard, A. Szasz, A. Sztein, M. Taylor, J. P. Thiruraman, D. Thor, D. Timucin, E. Tomita, A. Torres, M. M. Torunbalci, H. Tran, A. Vaishnav, J. Vargas, S. Vdovichev, G. Vidal, C. V. Heidweiller, M. Voorhees, S. Waltman, J. Waltz, S. X. Wang, B. Ware, J. D. Watson, Y. Wei, T. Weidel, T. White, K. Wong, B. W. K. Woo, C. J. Wood, M. Woodson, C. Xing, Z. J. Yao, P. Yeh, J. Yoo, E. Young, G. Young, A. Zalcman, R. Zhang, Y. Zhang, N. Zhu, N. Zobrist, Z. Zou, G. Bortoli, S. Boixo, J. Chen, Y. Chen, M. Devoret, M. Hansen, C. Jones, J. Kelly, P. Kohli, A. Korotkov, E. Lucero, J. Manyika, Y. Matias, A. Megrant, H. Neven, W. D. Oliver, G. Ramachandran, R. Babbush, V. Smelyanskiy, P. Roushan, D. Kafri, R. Sarpong, D. W. Berry, C. Ramanathan, X. Mi, C. Bengs, A. Ajoy, Z. K. Minev, N. C. Rubin, T. E. O'Brien, Quantum computation of molecular geometry via many-body nuclear spin echoes. arXiv arXiv:2510.19550 [Preprint] (2025). https://doi.org/10.48550/arXiv.2510.19550.

25. M. H. Levitt, Symmetry constraints on spin dynamics: Application to hyperpolarized NMR. *J. Magn. Reson.* **262**, 91–99 (2016).